\documentclass[aps,prb,twocolumn,superscriptaddress]{revtex4}
\usepackage{graphicx}
\usepackage{bm}
\begin{document}

\title{Field-induced gap in ordered Heisenberg  antiferromagnets}
\author{J.-B.~Fouet}
\affiliation{Institut Romand de Recherche Num\'erique en Physique des 
Mat\'eriaux (IRRMA), PPH-Ecublens, CH-1015 Lausanne, Switzerland} 
\author{O.~Tchernyshyov}
\affiliation{Department of Physics and Astronomy, Johns Hopkins University, 3400 N.
Charles St., Baltimore, MD 21218, USA}
\author{F.~Mila}
\affiliation{
Institute of Theoretical Physics, 
Ecole Polytechnique F\'ed\'erale de Lausanne, 
CH-1015 Lausanne, Switzerland}
\date{\today}
\begin{abstract}

Heisenberg antiferromagnets in a strong uniform magnetic field $H$ are
expected to exhibit a gapless phase with a global O(2) symmetry.  In
many real magnets, a small energy gap is induced by additional
interactions that can be viewed as a staggered transverse magnetic
field $h = c H$, where $c$ is a small proportionality constant.  We
study the effects of such a perturbation, particularly for magnets
with long-range order, by using several complimentary approaches:
numerical diagonalizations of a model with long-range interactions,
classical equations of motion, and scaling arguments.  In an ordered
state at zero temperature, the energy gap at first grows as
$(cH)^{1/2}$ and then may dip to a smaller value, of order
$(cH)^{2/3}$, at the quantum critical point separating the ``gapless''
phase from the gapped state with saturated magnetization.  In one
spatial dimension, the latter exponent changes to 4/5.
\end{abstract}

\maketitle

\section{Introduction}
The investigation of quantum antiferromagnets in strong magnetic fields is
currently a very active field of research.  Several remarkable
properties have been observed, such as the closing of the energy gap
in spin-1 chains\cite{zheludev} and spin ladders\cite{chaboussant} and
observations of magnetization plateaux in frustrated
magnets.\cite{kageyama,kodama} While the broad features of these
models can be explained in the framework of the Heisenberg model in a
uniform magnetic field, a closer examination of experimental data
reveals deviations from theoretically predicted behavior.  In
particular, the supposedly gapless phases actually possess small
energy gaps, which can only be explained by the presence of
anisotropic interactions.

For instance, Dender {\em et al.}\cite{D97} discovered that an applied
magnetic field ${\bf H}$ induces in spin-1/2 Heisenberg chains a gap
$\Delta \propto H^{0.65(3)}$.  Oshikawa and Affleck\cite{O97}
ascribed the gap to a staggered transverse field arising from the
staggering of the $g$-tensor or of the Dzyaloshinskii-Moriya (DM)
interaction.  They suggested the effective Hamiltonian
\begin{equation}
{\cal H} = \sum_{n} [J {\bf S}_n \cdot {\bf S}_{n+1}
- H S_n^{z}
- h \, (-1)^n S_n^{x} ],
\label{Ham}
\end{equation}
where $h \propto H$.  The transverse field ${\bf h}$ creates a spin
gap $\Delta \propto h^{2/3} \propto H^{2/3}$, in agreement with the
experiment and density-matrix renormalization group (DMRG)
calculations.\cite{L02,Zhao03}

An extension of these results to higher dimension is still at a
preliminary stage.  Sato and Oshikawa\cite{Sato04} 
studied a model of
interacting chains with a staggered field induced by the DM
interactions and found that the gap scales as $\Delta \propto H^{1/2}$
in a weak field.  The recent discovery\cite{kageyama} of the
Shastry--Sutherland antiferromagnet ${\rm SrCu_2(BO_3)_2}$ calls for
further studies in two dimensions.  In particular, the presence of a
staggered magnetization in the low-field NMR signal was ascribed to a
staggering of the $g$-tensor and of the DM interaction.\cite{kodama2}
The direct numerical investigation of the relevant microscopic models
does not seem to be possible at the moment.  Indeed, to get a reliable estimate of the
small gap induced by anisotropic interactions requires to reach 
sizes such that the finite-size gaps, typically of order $J/N$ for $N$
sites, are much smaller than the physical gap. This
is possible in 1D with the help of the DMRG algorithm, which by now
routinely allows one to study systems with 200 sites or more, 
but not in 2D. 

In this paper, we present the first systematic study of this problem
in the context of the Lieb-Mattis model,\cite{LM62} wherein every spin
of one sublattice is coupled equally to all spins of the other. This
model is expected to provide a fair description of the long-wavelength
properties of bipartite Heisenberg antiferromagnets with long-range
N\'eel order in the ground state.  Therefore it should be relevant for
magnets in 2 and 3 dimensions.

We pay particular attention to the high-field regime where the uniform
magnetization saturates.  In the absence of a staggered field, the
saturation occurs at a critical point $H = H_c$ separating a gapless
ordered phase with a spontaneously broken rotational O(2) symmetry
from a gapped phase with fully polarized spins.  The behavior of the
transverse-field spin gap near the critical field $H_c$ is an
important problem in view of its relevance to a number of experimental
systems.  Unlike the weak-field regime, for which analytical results
have been obtained, the saturation region has so far been studied only
numerically.\cite{L02,Zhao03} Curiously, the numerical data reveal a
nonmonotonic dependence $\Delta(H)$, with a pronounced minimum near
the saturation field $H_c$.  A similar effect was noticed earlier by
Sakai and Shiba in their numerical analysis of spin-1
chains.\cite{Sakai94}

\section{The Lieb--Mattis model}
In this section, we concentrate on the Lieb-Mattis model describing a
Heisenberg antiferromagnet on a bipartite lattice in which every spin
of one sublattice is coupled through antiferromagnetic exchange to
{\it all} sites of the other sublattice.  Its Hamiltonian describes
spins of length $S$ in a uniform magnetic field ${\bf H} = (0,0,H)$
and a staggered field ${\bf h} = (h,0,0)$ perpendicular to it:

\begin{eqnarray}
{\cal H} &=& (J/N)\sum_{\bf r \in A}\sum_{\bf r' \in B}
{\bf S}_{\bf r} \cdot {\bf S}_{\bf r'}
\nonumber\\
&& - \sum_{\bf r \in A} ({\bf H} + {\bf h}) \cdot {\bf S}_{\bf r} 
- \sum_{\bf r \in B} ({\bf H} - {\bf h}) \cdot {\bf S}_{\bf r}
\label{Ham-lieb}\\
&=& (J/N) \, {\bf S}_A \cdot {\bf S}_B - H(S_A^z+S_B^z) - h(S_A^x-S_B^x),
\nonumber
\end{eqnarray}
where ${\bf S}_A$ and ${\bf S}_B$ are the total spins of sublattices A
and B.  The exchange constant $J$ is normalized by the total number of
sites $N$ to make the energy of the model an extensive quantity
${\mathcal O}(N)$.  To reflect the induced origin of the staggered
field ${\bf h}$, we will set $h = c H$, where the constant $c \ll 1$
is a property of the system.

\begin{figure}
\centerline{\includegraphics[width=\columnwidth]{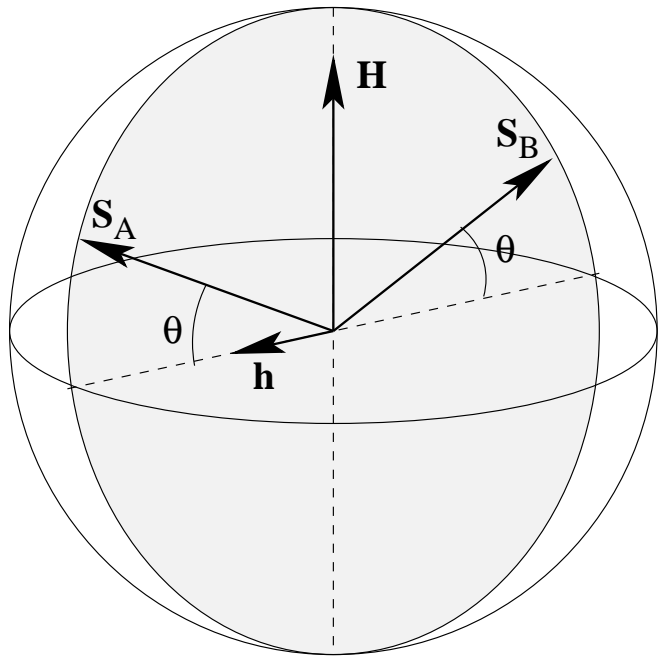}}
\caption{A classical ground state.  The sublattice spins reside in the
plane formed by the vectors ${\bf H}$ and ${\bf h}$.}
\label{fig-canted}
\end{figure}

In zero transverse field, the Hamiltonian (\ref{Ham-lieb}) has an O(2)
rotational symmetry and is readily diagonalized.  The energy levels
are expressed in terms of the total spins of the sublattices $S_A$ and
$S_B$, the total spin $S_{\rm tot}$ and its projection $S_{\rm tot}^z$
on the direction of the uniform field ${\bf H}$:
\begin{eqnarray}
E_0 &=& (J/2N) S_{\rm tot}(S_{\rm tot}+1) - H S_{\rm tot}^z
\nonumber\\
&-& (J/2N) [S_A(S_A+1) + S_B (S_B+1)].
\end{eqnarray}
In the ground state the sublattices are fully polarized, $S_A = S_B =
NS/2$, and the total angular momentum points along ${\bf H}$: $S_{\rm
tot}^{z} = S_{\rm tot}$.  Excitations reducing sublattice
magnetizations have an energy gap ${\mathcal O}(JS)$.  Below
saturation, $H \leq H_c = JS$, low-energy excited states are obtained
by changing the quantum numbers $S_{\rm tot}^{z} = S_{\rm tot}$ from
their ground-state values.  These excitations have energy ${\mathcal
O}(JS/N)$ and form a continuum in the limit $N \to \infty$.  The
gapless excitations are caused by spontaneous breaking of the O(2)
symmetry.  Indeed, for $N \to \infty$ the sublattice spins ${\bf S}_A$
and ${\bf S}_B$ become classical vectors with well-defined directions
in space (Fig.~\ref{fig-canted}).  In the absence of the transverse
staggered field ${\bf h}$, the sublattice moments can be freely
rotated about ${\bf H}$.  The low-energy excitations thus correspond
to a slow precession of ${\bf S}_A$ and ${\bf S}_B$ when the angle
$\theta$ deviates slightly from the equilibrium value $\theta =
\arcsin{(H/JS)}$.

Adding the transverse field ${\bf h}$ breaks the O(2) symmetry
explicitly and violates conservation of both $S_{\rm tot}$ and $S_{\rm
tot}^z$.  Nonetheless, lengths of the sublattice spins $S_A$
and $S_B$ are still good quantum numbers.  This fact greatly
simplifies the numerical diagonalization of the Hamiltonian
(\ref{Ham-lieb}).

\subsection{Exact diagonalizations}

Because the ground state and all low-lying excitations of the model
with ${\bf h} = 0$ are in the sector $S_A = S_B = N S/2$, we will
restrict our analysis of the gap to that sector. The size of this
subspace for $N$ spins equals $(NS+1)^2$, which is smaller than the
size of the total Hilbert space, $(2S+1)^N$.  This enables us to treat
systems with rather large numbers of spins. In the following, we
present results for 2000 spins $S = 1/2$, a size clearly beyond the
scope of exact diagonalizations of other Heisenberg models without
additional conserved quantities.

\begin{figure}
\centerline{\includegraphics[angle=-90,width=\columnwidth]{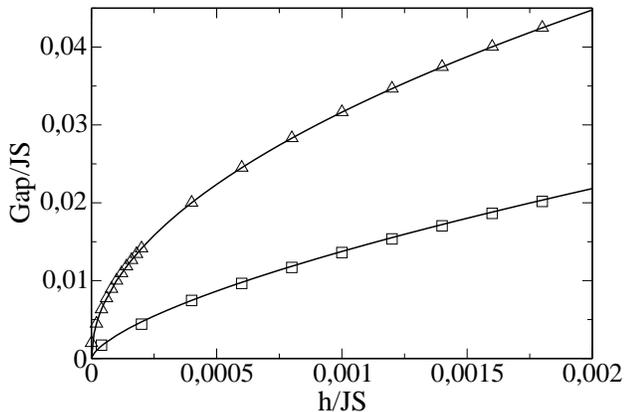}}
\caption{The spin gap $\Delta$ as a function of the transverse field
$h$ for two values of the uniform field: $H=0$ (triangles) and $H =
H_c = JS$ (squares) for $N=2000$. The curves are the results of the
spin-wave analysis.}
\label{fig-gapH=1_0}
\end{figure}

We first calculate the gap as a function of $h$ in the absence of a
uniform field. The results are plotted in Fig.\ref{fig-gapH=1_0}.  By
fitting the data at low field $h\to 0$, we determined that the gap
vanishes as $\Delta \sim \sqrt{JSh}$, precisely as found by Oshikawa
and Affleck in the approximation of noninteracting magnons.\cite{O97}
This result is most easily understood by computing the precession
frequency of sublattice magnetizations at the classical level, as we
do in the next section.  The result of this calculation is plotted as
a solid line in Fig.~\ref{fig-gapH=1_0}. An excellent agreement shows
that size effects are already negligible for $N = 2000$ (with the
exception of a finite gap at $h=0$).

\begin{figure}
\centerline{\includegraphics[angle=-90,width=\columnwidth]{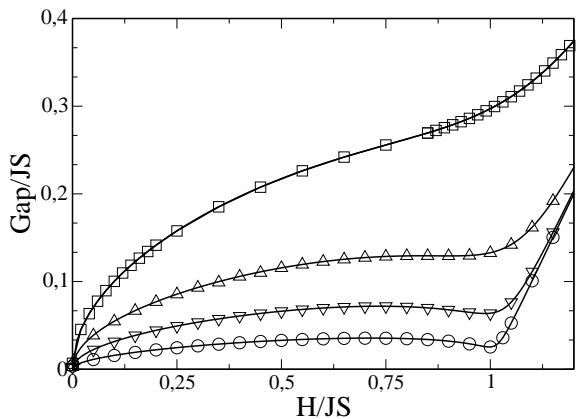}}
\caption{The spin gap as a function of $H$ at $h= c H$ for several
value of the proportionality constant: $c=0.1$ (squares), 0.03
(triangles up), 0.01 (triangles down) and 0.0025 (circles). The gap
has a local minimum for $c \leq 0.03$. Lines are the results of the
spin-wave analysis.}
\label{gapcomp}
\end{figure}

Next we consider the case where the staggered field is proportional to
the uniform field: $h = c H$. Typical results obtained for various
values of $c$ are plotted in Fig. \ref{gapcomp}.  As $H \to 0$,
$\Delta \sim H^{1/2}$ with numerical precision. The behaviour at
higher fields depends on the value of the proportionality constant
$c$.  If it is small enough, $c < 0.03 \pm 0.005$, the gap exhibits a
local minimum close to---and slightly below---the saturation field.
Such a dip has been observed previously in numerical studies of 1D
models,\cite{Zhao03} but no explanation has been given so far.  To get
further insight, we have kept the uniform field at its saturation
value $H_c = JS$ and calculated the gap as a function of $h$. The
results are plotted in Fig.\ref{fig-gapH=1_0}.  They are consistent
with a power-law scaling $\Delta \propto h^{2/3}$.  We will return to
this phenomenon in the next section.

\begin{figure}
\centerline{\includegraphics[angle=-90,width=\columnwidth]{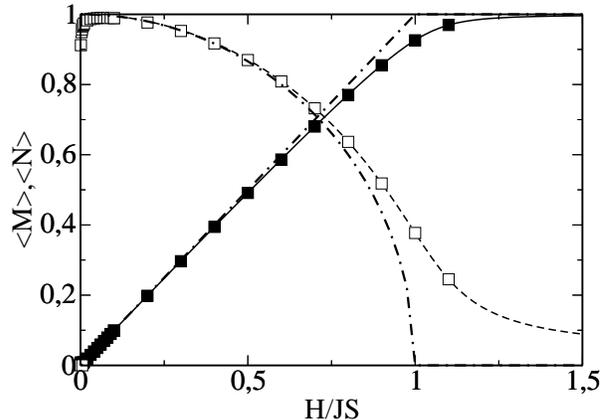}}
\caption{Uniform (filled squares) and staggered (open squares)
magnetizations for $N=2000$ as a function of $H$ for $c = 0.03$.
Solid and dashed lines are results of the spin-wave
analysis. Dash-dotted lines show the behavior for ${\bf h} = 0$.}
\label{fig-magdiag}
\end{figure}

Finally, we have calculated the uniform and staggered magnetizations
as functions of $H$ for several values of $c$ (see
Fig.~\ref{fig-magdiag}).  As expected, the introduction of the
symmetry-breaking staggered field removes the critical behavior near
$H = H_c = JS$.

\subsection{Semi-classical analysis}
\label{sec-anal}

The results of the exact diagonalizations can be understood in the
framework of the classical equations of motion for the sublattice
moments $ {\bf S_A}$ and ${\bf S_B}$.  The frequencies of classical
spin waves are identical to the energies of magnons obtained in the
lowest-order $1/S$ expansion, one of the methods employed by Oshikawa
{\em et al.}\cite{O97,Sato04,A99} Because the magnon gap is lowest at
zero wavevector, we specialize to uniform solutions, and consider
precession of sublattice spins ${\bf S}_A$ and ${\bf S}_B$ at maximal
length $NS/2$.

The equilibrium canting angle $\theta$ (Fig.~\ref{fig-canted}) is
determined by minimization of the classical energy:
\begin{equation}
JS^2 \sin \theta \cos \theta +hS \sin \theta-HS \cos\theta =0
\label{Eq-min}
\end{equation}
In the absence of the transverse field, $\theta = \arcsin{(H/H_c)}$
for $H \leq H_c = JS$ and $\pi/2$ for $H>H_c$.  For $h \neq 0$, the
kink in $\theta(H)$ is smoothed out.

The equations of motion for the sublattice moments $ {\bf S_A}$ and
${\bf S_B}$ are:
\begin{eqnarray}
\hbar \dot{\bf S}_A
&=& \left(J{\bf S}_B/N - {\bf H} - {\bf h}\right)\times{\bf S}_A, 
\nonumber\\
\hbar  \dot{\bf S}_B
&=& \left(J{\bf S}_A/N - {\bf H} + {\bf h}\right)\times{\bf S}_B.
\label{Eq-motion}
\end{eqnarray} 
We rewrite these equations in terms of the uniform and staggered
magnetizations ${\bf m} = ({\bf S}_A+{\bf S}_B)/N$ and ${\bf n} =
({\bf S}_A-{\bf S}_B)/N$ and linearize them in small deviations
$\delta {\bf m}$ and $\delta {\bf n}$ from the equilibrium values:
\begin{eqnarray}
\hbar \, \delta \dot{\bf m} 
&=& -{\bf H} \times \delta{\bf m} - {\bf h} \times \delta{\bf n}, 
\nonumber\\
\hbar \, \delta \dot{\bf n}
&=& -(J {\bf n} + {\bf h}) \times \delta{\bf m}
+ (J {\bf m} - {\bf H}) \times \delta{\bf n}.
\end{eqnarray}
The equilibrium values ${\bf m}$ and ${\bf n}$ have
length $S\cos \theta$ and $S \sin \theta$ respectively, and
the resulting precession frequencies are 
\begin{eqnarray}
\hbar\omega_{+}&=&\sqrt{H^2+h(JS \cos \theta+h)},
\nonumber\\
\hbar\omega_{-}&=&\sqrt{(H-JS\sin \theta)^2+h(JS\cos \theta+h)}.
\label{eigenfreq}
\end{eqnarray}
The slower mode involves the transverse components of ${\bf n}$ and the
longitudinal component of ${\bf m}$.  Its frequency determines the
energy gap: $\Delta = \hbar \omega_-$.
Figure~\ref{gapcomp} shows an essentially perfect agreement between
the classical analysis and the numerical diagonalization.  This is not
surprising: because the Lieb--Mattis model has infinite-range
interactions, the mean-field solution (\ref{Eq-min}) becomes exact in
the thermodynamic limit.

Next we explain the salient features seen in the dependence
$\Delta(H)$ (Fig.~\ref{gapcomp}), namely the initial increase $\Delta
\propto H^{1/2}$ and a dip around the saturation field $H_c = JS$ for
a small enough $c$.

\subsubsection{Weak uniform field: $H \ll JS$}
\label{sec-anal-weak}

Well below saturation, $H \ll H_c = JS$, the dominant effect is the
breaking of the axial symmetry by the transverse field ${\bf h}$.  To
estimate the resulting energy gap, we neglect the longitudinal field
${\bf H}$ and thus obtain the secular equation
\begin{equation}
-\hbar^2 \omega^2 \delta {\bf m} 
= {\bf h} \times [(J{\bf n} + {\bf h}) \times \delta {\bf m}].
\end{equation}
It yields the energy gap
\begin{equation}
\Delta = \hbar \omega = \sqrt{h(Jn+h)}.
\label{eq-weak-H}
\end{equation}
Spontaneous breaking of the axial symmetry implies that the staggered
magnetization ${\bf n}$ attains the maximal length $n = S$ for an
arbitrarily weak staggered field ${\bf h}$.  Therefore
\begin{equation}
\Delta \sim (JSh)^{1/2}
\label{eq-gap-weak}
\end{equation}
as $h \to 0$.  Recalling that $h = c H$ we obtain the gap $\Delta \sim
\sqrt{JScH}$ at low fields $H$, in accordance with
Eq.~(\ref{eigenfreq}).

\subsubsection{Uniform field at saturation: $H = JS$}
\label{sec-anal-sat}

For small enough $c$, the gap $\Delta(H)$ has a minimum near the
saturation field $H_c = JS$.  Its origin can be traced to a reduced
response of the spins to the transverse staggered field ${\bf h}$ at
saturation.  

Inspection of the equations of motion (\ref{Eq-motion}) in the absence
of the staggered field shows that the spin precession can be separated
into a fast mode with $\hbar\omega_+ = JS$ and a slow mode with
$\hbar\omega_- = 0$.  This separation of scales still works in the
presence of a weak transverse field.  
In contrast to the
phase with a spontaneously broken axial symmetry, the staggered
magnetization $n$ now vanishes as $h \to 0$.  With the aid of
Eq.~(\ref{Eq-min}), we obtain $n = S \cos{\theta} \approx
(2h/JS)^{1/3} S$ for a small $h$. Hence both terms in the expression
of $\hbar\omega_-$
in Eq.~(\ref{eigenfreq}) are of the same order since
$H-JS \sin \theta \approx (1/2)(2h/JS)^{2/3}$, 
 which yields the energy gap at $H = JS$:
\begin{equation}
\Delta \sim 3^{1/2}2^{-1/3}(JS)^{1/3}h^{2/3}
\label{eq-gap-sat}
\end{equation}

Comparison of Eqs.~(\ref{eq-gap-weak}) and (\ref{eq-gap-sat}) for a
staggered field $h = c H$ shows that the gap scales as $c^{1/2}$ at
weak fields $H$ but becomes of order $c^{2/3}$ at saturation.  The
latter is smaller (for a substantially small proportionality constant
$c$).  Therefore the initial increase of $\Delta$ with $H$ will be
followed by a dip at $H \approx JS$, provided that $c$ is small enough
(Fig.~\ref{gapcomp}).  The dip disappears when $c$ exceeds the
critical value $ c_c = 0.03126(6)$.

\section{Beyond the Lieb--Mattis model}

The results presented in this paper, most importantly the dependence
of the spin gap on the applied field $H$, were obtained for the
Lieb--Mattis model.  The special form of its Hamiltonian
(\ref{Ham-lieb}) has two important advantages.  First, the existence
of conserved quantities $S_A$ and $S_B$ enabled us to determine the
low-energy spectrum numerically for very large systems (up to $N =
2000$ spins).  Second, the infinite range of spin interactions
justified the use of a mean-field approximation, allowing us to derive
the low-energy spectrum and explain the observed features.  At the
same time, one must use caution in drawing conclusions for real
magnets on the basis of the results obtained in an infinite-range
model.  In this section, we discuss implications of our findings for
more realistic models---such as Eq.~(\ref{Ham})---in $d = 1$, 2, and 3
dimensions.

The problem of a generic Heisenberg antiferromagnet with two
sublattices in crossed uniform and staggered fields does not have an
exact solution.  Nonetheless, the considerations advanced in Section
\ref{sec-anal} can be extended to the general case of an
antiferromagnet with or without long-range N\'eel order.  To achieve
this goal, we use a field-theoretic approach, as was done previously
by Oshikawa and Affleck\cite{O97} in their work on the Heisenberg
chain.

\subsection{Effective models}

Let us identify the quantum field theory appropriate for the gapless
state (${\bf h} = 0$) and in its vicinity ($h \ll J$).  The uniform
field ${\bf H} \neq 0$ breaks the rotational symmetry O(3) down to
O(2) confining the staggered magnetization ${\bf n} = n(\cos{\phi},
\sin{\phi}, 0)$ to the $xy$ plane.  In $d \geq 2$
dimensions,\cite{remark1} the staggered magnetization acquires a
nonzero expectation value breaking the O(2) symmetry.  The low-energy
degrees of freedom are long-wavelength fluctuations of the direction
of ${\bf n}$ in the $xy$ plane, which can be thought of as a Bose
condensate\cite{Affleck} $n_x + i n_y = n e^{i\phi}$.  The spin waves
are fluctuations of its phase $\phi$; fluctuations of the amplitude
$n$ are gapped and for this reason can be neglected.  Thus one obtains
an effective Lagrangian for the low-energy excitations:
\begin{equation}
{\mathcal L} = \frac{\rho_s}{2} 
(\dot{\phi}^2/s^2 - |\nabla \phi|^2)
+ h n \cos{\phi}.
\label{eq-L-eff}
\end{equation}
Here $\rho_s > 0$ is a spin stiffness, and $s$ is the magnon velocity.
The last term $h n \cos{\phi}$, describing the coupling to the
staggered transverse field, breaks the residual O(2) $\equiv$ U(1)
symmetry and induces an energy gap.  In $d = 1$ dimension,
long-wavelength phase fluctuations destroy the condensate, $\langle n
e^{i\phi} \rangle = 0$; however, the low-energy theory
(\ref{eq-L-eff}) is still applicable.\cite{Affleck}

When the uniform field ${\bf H}$ reaches a critical magnitude $H_c$,
the antiferromagnet enters a polarized phase where all spins point
along the field direction, as in a ferromagnet.  For $H \geq H_c$, the
equilibrium magnitude of the staggered magetization $n$ vanishes and
the effective field theory (\ref{eq-L-eff}) no longer applies:
amplitude fluctuations become soft at $H = H_c$.  Above the critical
field, the magnon spectrum acquires a gap $\delta \propto H-H_c$ and
the energy dispersion switches from linear to quadratic, as in a
ferromagnet: $\epsilon_{\bf k} = \delta + k^2/2m$.  In the boson
language, $H=H_c$ can be viewed as the point of Bose condensation for
magnons.\cite{Affleck}  (There is, in fact, an exact mapping between
spins and hard-core bosons\cite{Friedberg} in the case of $S=1/2$.)
The low-energy effective theory describing universal properties of
interacting bosons near the condensation point has been discussed by
Fisher {\em et al.}\cite{Fisher}  It has the Lagrangian 
\begin{equation}
{\mathcal L} = -i \Phi^* \frac{\partial \Phi}{\partial t} -
\frac{|\nabla \Phi|^2}{2m} - \delta |\Phi|^2 - \lambda |\Phi|^4 +
h(\Phi^* + \Phi),
\label{eq-L-eff-bosons}
\end{equation}
where $\Phi = n_x + i n_y$. When $\delta < 0$, the bosons condense,
$\Phi = n e^{i\phi}$, and the effective field theory reduces to 
that of phase fluctuations (\ref{eq-L-eff}).  

Next we discuss the properties of the effective models below and at
the condensation point, with particular emphasis on the energy gap
induced by the staggered transverse field ${\bf h}$.

\subsection{Weak uniform field: $H \ll H_c$}

Well below the condensation point, phase fluctuations are the dominant
excitations.  The effective field theory is given by the Lagrangian
shown in Eq.~(\ref{eq-L-eff}).  Analytical continuation to imaginary
times $t = -i s\tau$ yields the classical XY model in the ordered
phase in $d+1$ dimensions, whose properties are well
known.\cite{Chaikin} In particular, the symmetry breaking field $h$
creates a finite correlation length $\xi \sim (h/\rho_s)^{-1/2}$.
This translates into an energy gap $\Delta \sim c(h/\rho_s)^{1/2}$ in
the quantum case.  The scaling of the gap is the same as in the
Lieb--Mattis model (\ref{eq-gap-weak}).

In $d = 1$, the long-range order is absent.  The ground state of the
quantum model in zero transverse field corresponds to the
Kosterlitz-Thouless phase with power-law correlations $\langle
e^{i\phi({\bf r})} e^{i\phi(0)}\rangle \sim C/r^\eta$ with a
nonuniversal exponent $\eta > 0$.  By the standard scaling
argument,the gap opens as $\Delta \propto h^{1/(2-\eta/2)}$.  For the
$S=1/2$ Heisenberg chain in the limit of zero uniform field, Oshikawa
and Affleck\cite{O97} find that $\Delta \propto h^{2/3}$.

\subsection{Uniform field at saturation: $H = H_c$}

Near the condensation point $H = H_c$ one must use the more complete
theory (\ref{eq-L-eff-bosons}) taking into account fluctuations of the
condensate magnitude $|\Phi| = n$.  This critical theory has been
analyzed by Fisher {\em et al.} in the context of the
superfluid--insulator transition.\cite{Fisher} Analytical continuation
to imaginary time $t = -i \tau$ yields a classical field theory in
$d+1$ dimensions; the singular part of the free energy density has a
scaling form
\begin{equation}
f_s(\delta, h) \sim \delta^{-(d+z)/y_\delta} X(h \, \delta^{-y_h/y_\delta}), 
\end{equation}
where $y_\delta$ and $y_h$ are the RG eigenvalues of the ``uniform
field'' $\delta$ and the staggered field $h$; $z$ is the dynamical
critical exponent.  

At the saturation field $H_c$ ($\delta=0$), the spin gap scales as a
power of the transverse field: $\Delta \propto h^{z/y_h}$.  The
exponents $z$ and $y_h$ are most readily obtained from the transverse
spin correlation function, whose long-wavelength part has the scaling
form
\begin{equation}
\langle \Phi^*({\bf r},\tau) \Phi(0,0)\rangle
\sim  \frac{Y(r \delta^{1/y_\delta}, \tau \delta^{z/y_\delta})}
{\tau^{2(d+z-y_h)/z}}. 
\end{equation}
The transverse spin correlations can be computed exactly in the polarized
phase ($H \geq H_c$), where the ground state is trivial and elementary
excitations are magnons with a quadratic dispersion:
\begin{equation}
\langle S^{+}({\bf r}, t) S^{-}(0, 0) \rangle
\propto
t^{-d/2} 
\exp{\left( -i \delta t - \frac{i m r^2}{2t}\right)},
\label{eq-corr-S}
\end{equation}
where $m$ is the magnon mass.  Hence $z = y_\delta = 2$ and $y_h = 2 +
d/2$.  The $\delta = h = 0$ fixed point is Gaussian in any
dimension.\cite{Fisher} The upper critical dimension is $d_c = 2$.

These considerations yield the following dependence of the spin gap at
the critical value of the uniform field $H = H_c$.  In $d = 1$,
$\Delta \propto h^{4/5}$.  In $d \geq d_c = 2$, the critical exponents
revert to their mean-field values, so that $\Delta \propto h^{2/3}$,
as in the Lieb--Mattis model.  

\begin{figure}
\centerline{\includegraphics[width=0.95\columnwidth]{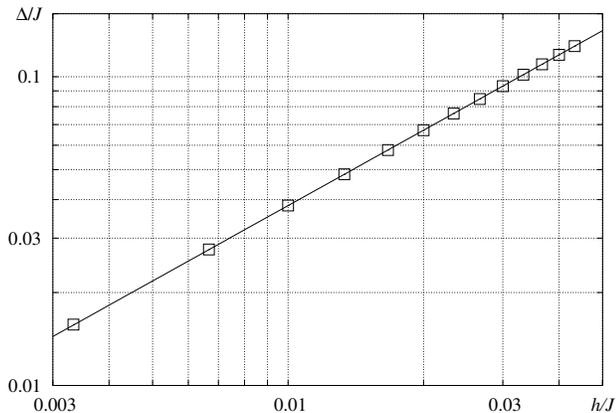}}
\caption{The energy gap $\Delta$ as a function of the staggered
transverse field $h$ in a $S=1/2$ Heisenberg chain of length $N=100$
in saturating uniform field $H = H_c = J$.  The squares are numerical
results obtained by the DMRG method.  The straight line is the
best-fitting power law $\Delta = a h^b$ with $b = 0.81$.}
\label{fig-gap-Hc}
\end{figure}

To test the validity of the field-theoretic argument, we have computed
numerically the energy gap $\Delta$ as a function of the staggered
transverse field $h$ in a $S=1/2$ Heisenberg chain.  Previous authors
have examined $\Delta(h)$ in the absence of the uniform field
$H$,\cite{L02,Zhao03} but not at the saturation point $H = H_c = J$.
We have employed the DMRG method for system sizes up to $N =
100$.\cite{Noak} The number of states kept was set at 150 for the
warm-up phase and 300 for the zip iteration.  With the exception of
the gapless point $h = 0$, size effects are rather weak.  In a chain
with $N = 100$ sites and $J = 1$ we find $\Delta = 1.60 h^{0.81}$
(Fig.~\ref{fig-gap-Hc}).  The numerical results are in reasonable
agreement with the predicted gap exponent of 4/5.

As expected on physical grounds, the energy gap $\Delta(h)$ opens more
slowly at the critical uniform field $H_c$ than in zero field.  If the
staggered field is proportional to the uniform one, $h = c H$, the gap
is a quantity ${\mathcal O}(c^{1/2})$ when $H$ is small and ${\mathcal
O}(c^{2/3})$ at saturation field $H_c$ (in $d \geq 2$).  Consequently,
for a small enough $c$, the gap will have a local minimum near $H =
H_c$, as in the Lieb--Mattis model.

\section{Conclusions}

The model of a Heisenberg antiferromagnet in a uniform magnetic field
$H$ perpendicular to a staggered magnetic fields $h = cH$ is relevant
to real magnets with a staggered $g$ tensor or staggered
Dzyaloshinskii--Moriya interaction.\cite{D97} We have determined the
low-energy spectrum of such a magnet with long-range order in the
model of Lieb and Mattis (\ref{Ham-lieb}).  The existence of conserved
quantities beyond the total spin allowed us to determine the spectrum
numerically in large systems (up to $N = 2000$ spins).  By utilizing
the infinite range of interactions in the model, we have also
reproduced the spectrum analytically and found essentially perfect
agreement with the numerical results.  

We have determined that the energy gap, caused by the breaking of the
O(2) spin-rotation symmetry by the induced staggered field, scales as
$(cH)^{1/2}$ at low uniform fields and is of order $(cH)^{2/3}$ at the
saturation field $H = H_c$.  The gap has a local minimum near $H_c$ if
$c$ is small enough ($c < 0.03126$).  Such a dip has been previously
observed in numerical studies of spin chains,\cite{L02,Zhao03} but has
not been explained.

Finally, we have presented scaling arguments establishing the same
power laws for the gap, and hence the existence of a saturation dip at
small $c$, to any Heisenberg antiferromagnet in magnetic field with
long-range order in $d \geq 2$ dimensions.  We note that a local
minimum of the spin gap near the saturation field has been observed in
${\rm SrCu_2(BO_3)_2}$, a Shastry--Sutherland magnet with staggered
$g$-factors and Dzyaloshinskii-Moriya interactions.\cite{kodama2}

\section*{Acknowledgements}

The authors thank C. Broholm and D.H. Reich for useful discussions and
R. Noak for providing his DMRG code and for assistance.  This work was
supported in part by the Swiss National Fund and by the US National
Science Foundation under the Grant No.~DMR-0348679.

\bibliographystyle{prsty}

\begin{thebibliography}{10}
\bibitem{Noak} We have used a computer code provided to us by
R.~Noak.
\end{thebibliography}

\end{document}